\documentclass[twocolumn,showpacs,preprintnumbers,amsmath,amssymb,superscriptaddress]{revtex4}

\usepackage{graphicx}% Include figure files
\usepackage{dcolumn}% Align table columns on decimal point
\usepackage{bm}% bold math

\begin{document}

%\preprint{APS/123-QED}

\title{Efficiency of Information Spreading in a population of diffusing agents}

\author{E. Agliari}
\affiliation{Dipartimento di Fisica, Universit\`a degli Studi di
Parma, Parco Area delle Scienze 7/A, 43100 Parma, Italy}
\author{R. Burioni}
\affiliation{Dipartimento di Fisica, Universit\`a degli Studi di
Parma, Parco Area delle Scienze 7/A, 43100 Parma, Italy}
\affiliation{INFN, Gruppo Collegato di Parma, Parco Area delle
Scienze 7/A, 43100 Parma, Italy}
\author{D. Cassi}
\affiliation{Dipartimento di Fisica, Universit\`a degli Studi di
Parma, Parco Area delle Scienze 7/A, 43100 Parma, Italy}
\affiliation{INFN, Gruppo Collegato di Parma, Parco Area delle
Scienze 7/A, 43100 Parma, Italy}
\author{F.M. Neri}
\affiliation{Dipartimento di Fisica, Universit\`a degli Studi di
Parma, Parco Area delle Scienze 7/A, 43100 Parma, Italy}

%\date{\today}

\begin{abstract}
We introduce a model for information spreading among a population of
$N$ agents diffusing on a square $L\times L$ lattice, starting from
an informed agent (Source). Information passing from informed to
unaware agents occurs whenever the relative distance is $\leq 1$.
Numerical simulations show that the time required for the
information to reach all agents scales as $N^{-\alpha}L^{\beta}$,
where $\alpha$ and $\beta$ are noninteger. A decay factor $z$ takes
into account the degeneration of information as it passes from one
agent to another; the final average degree of information of the
population, $\mathcal{I}_{av}(z)$, is thus history-dependent. We
find that the behavior of $\mathcal{I}_{av}(z)$ is non-monotonic
with respect to $N$ and $L$ and displays a set of minima. Part of
the results are recovered with analytical approximations.
\end{abstract}

\maketitle

\section{\label{sec:intro}Introduction}
The information spreading in a population constitutes an
attracting problem due to the emerging complex behavior and to the
great number of applications \cite{born,llas,moreno,hed}. The
propagation of information can be seen as a sequence of
interpersonal processes between the interacting agents making up
the system. In general, the population can be represented by a
graph where agents are nodes and links between them exist whenever
they interact with each other.

Authors, who previously investigated the diffusion of information
according to such a model, introduced different kinds of
interpersonal interaction, but almost all of them assumed a {\it
static} society \cite{moreno,guardiola,brajendra} (a notable
exception being that of Eubank {\it et al.} \cite{eubank}). In fact,
networks are usually built according to {\it a priori} rules, which
means that agents are fixed at their positions and can only interact
with their (predetermined) set of neighbors (the flow of information
between two agents is permanently open for linked pairs of agents
and permanently closed for non-linked pairs).

On the other hand, real systems are far from being static:
nowadays individuals are really dynamic and continuously come in
contact, and lose contact, with other people. Hence, the
interactions are rather instantaneous and time-dependent, and so
should be considered the links of the pertaining graph. The
network should be thought of as continuously evolving, adapting to
the new interpersonal circumstances.

Indeed, in sociology, where information spreading throughout a
population is a long-standing problem \cite{rapoport}, it is
widely accepted that processes of information transmission are far
from deterministic. Rather, they should incorporate some
stochastic elements arising, for example, from ``chance encounters
with informed individuals" \cite{allen}.

Sociologists also underline that, irrespective of the kind of
object to be transmitted, a realistic model should take into
account whether the object passed from one agent to another is
modified during the process \cite{borgatti}. Especially
information, which spreads by replication rather than
transference, is continuously revised while flowing throughout the
network. Degradation during transmission processes could reveal
important qualitative and quantitative effects, as some recent
works \cite{zhongzhu,lopez} started to point out.

This paper introduces a model that takes into account both the
issues discussed above, namely, a mobile society and information
changing during transmission. The model is based on a set of
random walkers meant as ``diffusing individuals": a population of
$N$ interacting agents embedded on a finite space is represented
by $N$ random walkers diffusing on a square $L \times L$ lattice.
We assume that two or more of them can interact if they are
sufficiently close to each other: as a result, a given agent has
no fixed position nor neighbors, but the set of agents it can
interact with is updated at each instant.

The information carried by an agent is a real (i.e., not boolean)
variable, whose value lies between 0 and 1. This (together with the
diffusive dynamics) is the main point that differentiates our model
from the susceptible-infected (SI) contact model of virus spreading
in epidemiological literature \cite{anderson}, where only two
status, susceptible and infected, are available to an agent. The
issue of information changing is dealt with by introducing a decay
constant $z\leq 1$, which measures the corruption experienced by the
piece of information when passing from an agent to another. We
assume $z$ to be universal: the more passages the information has
undergone before reaching an individual, the more altered it is with
respect to its original form.

\begin{figure*}[t]
\includegraphics[width=.9\textwidth]{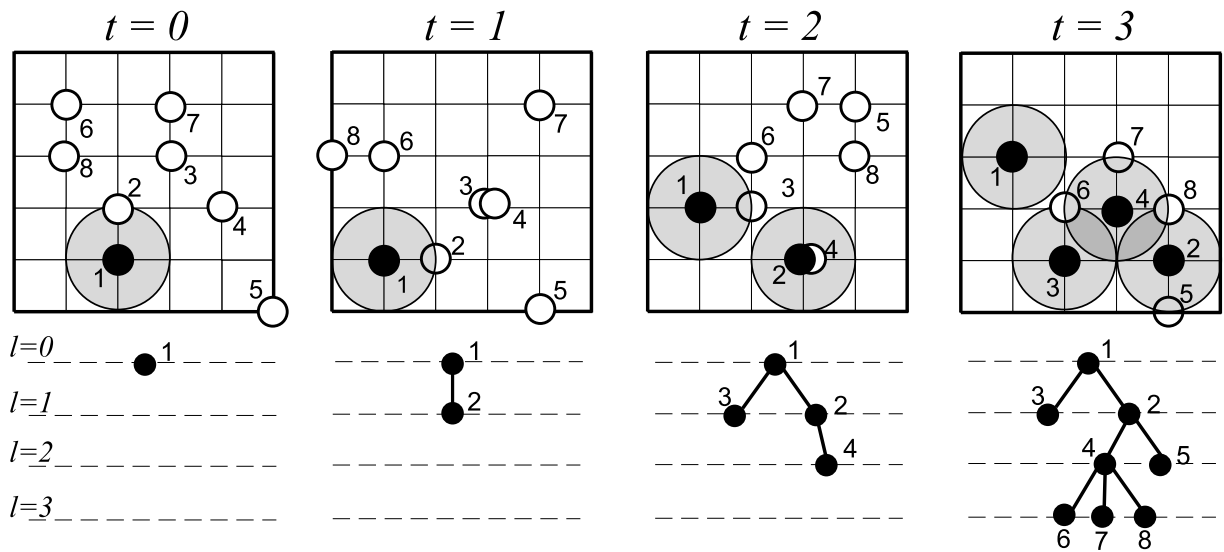}
\caption{\label{fig:example} Evolution of 8 agents on a $5\times 5$
lattice for $t$ from 0 to 3. For each $t$ the lattice is shown above
and the Information Tree is shown below. Informed agents are black
circles; unaware agents are white circles. A grey circle of radius 1
is drawn around every informed agent to represent its action (an
agent is in contact with another if it falls within this circle).
$\mathbf{t=0}$: the only informed agent is the Information Source
which carries information 1, so $n(0,0)=1$ and $n(0,l)=0$ for $l>0$.
$\mathbf{t=1}$: agent 1 passes information to agent 2; now
$n(1,0)=1$, $n(1,1)=1$; $\mathbf{t=2}$: agent 1 passes information
to agent 3 and agent 2 passes information to agent 4; $n(2,0)=1$,
$n(2,1)=2$, $n(3,1)=1$; $\mathbf{t=3}$: agent 2 passes information
to agent 5; agent 4 passes information to agents 6, 7, 8. Notice
that agent 6 is in contact with both 3 and 4; it chooses randomly to
get information from 4 (the same for agent 8). Now all agents have
been informed: for this simulation the Population-Awareness Time is
$\tau = 3$. The final information is
$\mathcal{I}(\tau)=\sum_{l=0}^{\tau}n(l,\,\tau)z^l=1+2\,z+2\,z^2+3\,z^3$.}
\end{figure*}

We study the time it takes for the piece of information to reach
every agent (Population-Awareness Time). We show that it depends
on $N$ and $L$ as a power-law, whose exponents are constant with
respect to system parameters. We also investigate the final
average (per agent) degree of information $\mathcal{I}_{av}(z)$.
We show that $\mathcal{I}_{av}(z)$ is not a monotonic function of
the density $\rho=\frac{N}{L^2},$ but displays minima for definite
values of $N$, $L$. This interesting result implies that there
does not exist a trivial direction where to tune the system
parameters $N$ and $L$ in order to make information spreading more
efficient.

In the following, we first describe our model
(Sec.~\ref{sec:model}), then we expose results obtained by means
of numerical simulations (Sec.~\ref{sec:NumRes}). Next,
Sec.~\ref{sec:Analytical} contains analytical results which corroborate
and highlight the former. Finally,
Sec.~\ref{sec:Conclusions} is devoted to our conclusions and
perspectives.

\section{\label{sec:model}The model}
$N$ random walkers (henceforth, agents) move on a square $L\times
L$ lattice with periodic boundary condition. At time $t=0$ the
agents are randomly distributed on the lattice. At each instant $t>0$
each agent jumps randomly to one of the four nearest-neighbor sites.
There are no excluded-volume effects: there can be more agents on the
same site; $\rho=N/L^2$ is the density of agents on the lattice.\\
Each agent $j$ carries a number $I_j$, $0\leq I_j \leq 1$,
representing information; an agent is called ``informed'' if
$I_j>0$ and ``unaware'' if $I_j=0$. At $t=0$ one agent, say agent
1, carries information 1 and is called the Information Source (or
simply the Source); the other $N-1$ agents are unaware. The aim of
the dynamics is to diffuse information from the Source to all
agents. \\Interaction between two agents $j$ and $k$ takes place
when i) one of them is informed and the other unaware, and ii) the
chemical distance between the two agents is $\leq 1$ (i.e., they
are either on the same site or on nearest-neighbor sites: we then
say that they are ``in contact''). By ``interaction'' we mean an
information passing from the informed agent, say $j$, to the
unaware one $k$ with a fixed decay constant $z$ ($0\leq z\leq 1$):
if $j$ carries information $I_j$, then $k$ becomes informed with
information $I_k=z\,\cdot I_i$. Once an agent has become informed,
it will never change nor lose its information (that is, informed
agents never interact). If an unaware agent comes in contact with
more informed agents at the same time, each carrying its own
information $I_j$, it will acquire the information of one of them
chosen at random (multiplied by $z$). The simulation stops at the
time $\tau$ when all the agents have become informed: we call this
the Population-Awareness Time (PAT).
\\
We define $n(t)$ the total number of informed agents at time $t$
($n(0)=1$; $n(\tau)=N$). As a result of our model, the information
carried by an agent is always a power of the decay constant $z^l$,
where $l$ is the number of passages from the Information Source to
the agent. We say that an informed agent belongs to level $l$ when
it has received information after $l$ passages from the
Information Source. We call $n(l,t)$ the number of agents
belonging to the $l$-th level at time $t$, or the population of
level $l$ at time $t$: $n(t)=\sum_{l=0}^{t}n(l,t)$. In Fig.
\ref{fig:example} we show as an example the evolution of $N=8$
agents on a $5\times 5$ lattice.
\\
We can envisage information passing by drawing an Information Tree with
$N$ nodes and $N-1$ links (fig. \ref{fig:example}): the agents are the
nodes of the tree, and a link is drawn between two agents when one passes
information to the other. An agent belongs to level $l$ if its distance
from the Source along the tree is $l$. The Information Tree evolves with time:
the tree at instant $t$ is a subtree of that at instant $t+1$.\\
At each instant $t$ we define the total information
\begin{equation}
\label{eq:tot_information}
\mathcal{I}(z,\,t)=\sum_{l=0}^{t}n(l,t)z^l;
\end{equation}
notice that it is the generating function of $n(t)$; consequently,
$$n(t) = \mathcal{I}(1,t).$$ We are interested in particular in
the final information
$$\mathcal{I}(z) =\mathcal{I}(z,\tau).$$
and in its average value per agent, $\mathcal{I}_{av}(z)=\mathcal{I}(z)/N$.

\section{\label{sec:NumRes}Numerical Results}

\begin{figure}[t]
\includegraphics[width=.45\textwidth]{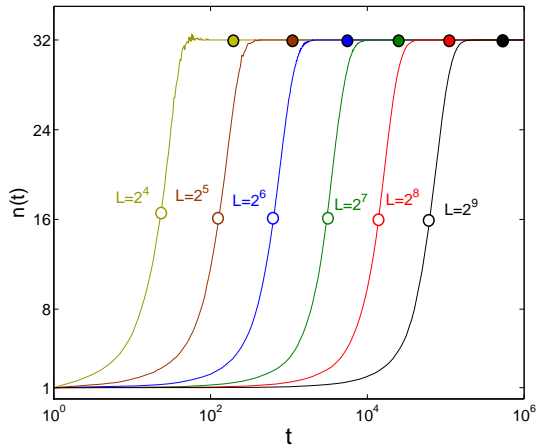}
\caption{\label{fig:time_evolution}  (Color online) Evolution of
$n(t)$ for a population of $N=32$ agents on six different lattices
of size $L=2^m$, $m=4,\ldots,9$. Full circles denote the
Population-Awareness Times, empty circles the Outbreak Times. }
\end{figure}

This section is divided into three parts. The first considers only $n(t)$,
the total informed population at time $t$, and the
results presented are independent of the population distribution on levels.
The second section takes into account the distribution on levels $n(l,t)$. The third
section deals with the final information $\mathcal{I}(z)$.
All the results are averaged over 500 different realizations
of the system.

\subsection{\label{sec:num} Level-independent results}

\begin{figure}[b]
\includegraphics[width=.45\textwidth]{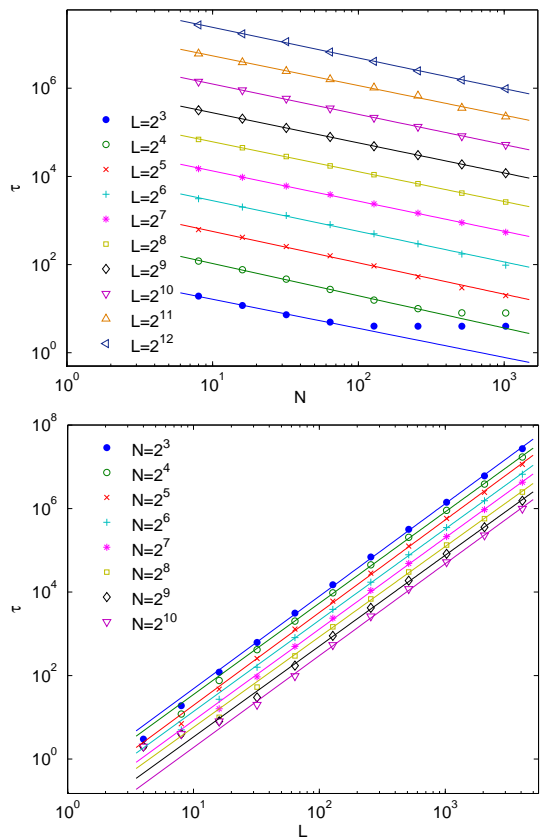}
\caption{\label{fig:PowLaw_NL}(Color online) Dependence of the
Population-Awareness Time $\tau$ on the number of agents $N$  and
the lattice size $L$. Top: Log-log scale plot of $\tau$ versus $N$;
different lattice-size values are shown with different symbols and
colors. For sufficiently small densities ($\rho \leq 1$), straight
lines represent the best fit according to Eq.~\ref{eq:PowLaw}.
Bottom: Log-log scale plot of $\tau$ versus $L$; different values of
the number of agents are shown with different symbols and colors.
Provided that the density $\rho$ is not large ($\rho \leq 1$), data
points lay on the curves given by Eq.~\ref{eq:PowLaw} which
represent the best fit. Error on data points is $<2 \%$.}
\end{figure}

Fig. \ref{fig:time_evolution} shows the typical time evolution of
$n(t)$, the number of aware people at time $t$, for fixed $N$ and
several different values of $L$. Due to the fact that, once
informed, an agent can not modify his status, $n(t)$ is a
monotonic increasing function. The curve is sigmoidal: $n(t)$
initially increases with an increasing growth rate $dn(t)/dt$. The
growth rate is maximum at the Outbreak Time $t_{out}$, when
usually $n(t_{out})\sim N/2$ (in Sec. \ref{sec:Analytical} we will
justify this fact in a low-density approximation). The growth rate
then begins to decrease; the evolution slows down and the curve
begins to saturate. The information reaches all the population at
the Population-Awareness Time $\tau$, that is the quantity that we
analyze here (roughly $\tau\sim 2t_{out}$, and this fact as well
will be justified in Sec. \ref{sec:Analytical}).

The Population-Awareness Time $\tau$ depends on the total number of
agents $N$ and on the size of the lattice $L$, as shown in
Fig.~\ref{fig:PowLaw_NL}. As long as the density is not large ($\rho
\leq 1$), data points are well fitted by power laws holding over a
wide range (though logarithmic corrections can not be ruled out):
\begin{eqnarray}
&& \tau \sim N^{-\alpha},\\
&& \tau \sim L^{\beta}.
\end{eqnarray}

The exponents $\alpha$ and $\beta$ are
constant by varying $L$ or $N$, respectively, so that we can
write:
\begin{equation}
\label{eq:PowLaw}
\tau\sim N^{-\alpha}\, L^{\beta}.
\end{equation}

The fitting of data with an asymptotic least-squares method yields
the following exponents:
\begin{equation}\label{eq:alpha_beta}
\alpha = 0.68 \pm 0.01 \hspace{3ex}   \beta = 2.22 \pm 0.03.
\end{equation}

\subsection{\label{sec:num_pop} Level-dependent results}

\begin{figure}[b]
\includegraphics[width=.45\textwidth]{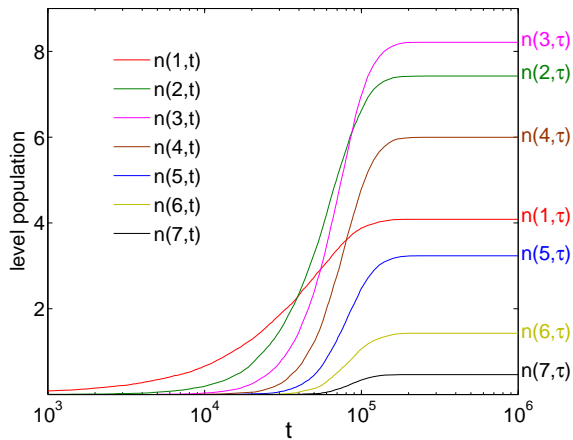}
\caption{\label{fig:levels} (Color online) Time evolution of level
populations $n(l,\,t)$ for $N=32$, $L=512$. }
\end{figure}

\begin{figure}[t]
\includegraphics[width=.5\textwidth]{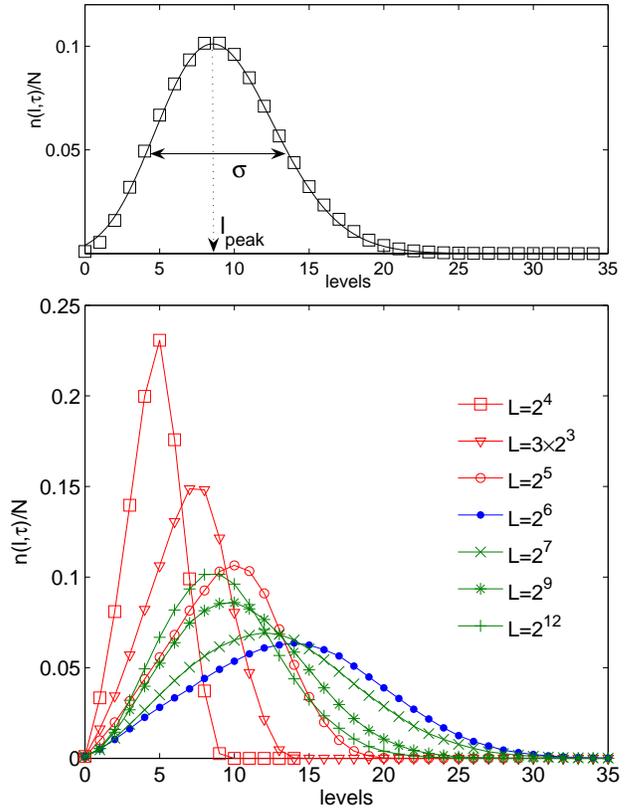}
\caption{\label{fig:level_distribution}  (Color online) Top: typical
population distribution on levels at $t=\tau$ for a low-density
system (N=1024, L=4096). Squares are experimental results; the line
is the result of data fitting according to Eq.
(\ref{eq:fitting_function}). Bottom: population distribution on
levels at $t=\tau$ for systems with $N=1024$ and $L$ between $2^4$
and $2^{12}$ (the lines are guides to the eye): the behavior of the
distribution is non-monotonic with respect to $L$. By increasing $L$
from small values, the curves first shift to the right and flatten
($L=16$, 24, 32). The rightmost, extremal curve corresponds to
$L=64$. Then, by increasing $L$ the curves shift back to the left
and sharpen ($L=128$, 512, 4096).}
\end{figure}

We now focus on the time evolution of $n(l,\,t)$, the population of
level $l$. Each population evolves in time with a sigmoidal law
(Fig. \ref{fig:levels}), with its own Outbreak Time and tending to a
final value $n(l,\,\tau)$.

The final distribution of agents on levels $n(l,\,\tau)$ as a
function of $l$ (Fig. \ref{fig:level_distribution}, top) has an
asymmetrical-bell shape, with a peak at position $l_{peak}$ and a
width $\sigma$, both depending on $N$ and $L$ (notice that only a
fraction of the $N$ available levels has a non-negligible
population). If $L$ is large enough (larger than $\tilde{L}$, see
below), the population distribution on levels is well fitted by the
$3$-parameter function
\begin{equation}\label{eq:fitting_function}
\frac{n(l,\tau)}{N}=A\,\frac{\left(\mathrm{log}N\right)^l}{\Gamma(B\cdot
l+C)},
\end{equation}
where $\Gamma(x)$ is the Euler gamma function, and the parameters
$A, B, C$ depend smoothly on $N$ and $L$. The fitting function is
a generalization of Eq. (\ref{eq:lev_distr_highdil}), the
distribution function of the low-density regime.

In Fig. \ref{fig:level_distribution}, bottom, we show how the
distribution $n(l,\,\tau)$ changes with $L$ for a fixed value
$N=1024$ and we introduce one of the most important results of this
paper. For $L$ small (hence for high density, $\rho\gg1$) the
distribution is very sharp and peaked on small values of $l$. As $L$
grows, the distribution shifts to higher values of $l$ and becomes
more and more spread ($l_{peak}$ and $\sigma$ grow). The extremal,
maximum-spread distribution is obtained for a value $L=\tilde{L}$
(for $N=1024$, $\tilde{L}\sim 64$): $l_{peak}$ and $\sigma$ are
maximum; the highest possible number of levels is occupied. As $L$
is increased beyond $\tilde{L}$, the curve begins to shift back to
smaller $l$s and to narrow; this process continues up to the
low-density regime ($\rho\ll1$). In general, $\tilde{L}$ depends on
$N$.

The same phenomenon occurs if we keep $L$ fixed and let $N$ vary. By
increasing $N$ from small, low-density values, the distribution
shifts to the right and spreads, up to an extremal form occurring
for $N=\tilde{N}$ (depending on $L$). Then, it shifts back and
narrows.

This behavior has strong consequences on the efficiency of
information spreading on the lattice, as we will see in the next
section.

\subsection{\label{sec:FinInfo} Degree of Information}
In this section we deal with the final degree of information at
the Population-Awareness Time, $\mathcal{I}(z) = \mathcal{I}(\tau,z)$ (in particular, with its
average value $\mathcal{I}_{av}(z)=\mathcal{I}(z)/N$), and
its dependence on $N$, $L$, and $z$. We remind (Eq. (\ref{eq:tot_information})) that
$\mathcal{I}(z)$ is the generating function of the final populations $n(l,\,\tau)$,
hence its value depends on the final distribution of the population on levels
analyzed in the previous paragraphs.

\begin{figure}[b]
\includegraphics[width=.5\textwidth]{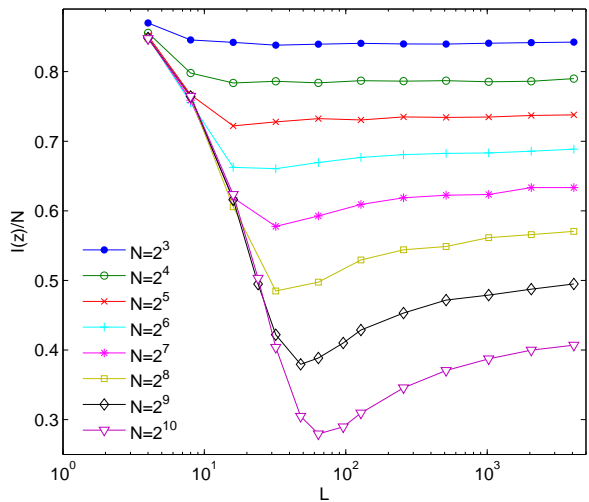}
\caption{\label{fig:Info} (Color online) Semilog scale plot of final
degree of information per agent
$\mathcal{I}_{av}(z)=\mathcal{I}(z)/N$ vs lattice size $L$. Several
values of $N$ are shown with different symbols and colors (lines are
guides to the eye), while the decay constant is fixed at $z=0.9$.
Notice the occurrence of minima at $\tilde{L}, \tilde{N}$, and that
$\tilde{L}$ is monotonically increasing with respect to $\tilde{N}$.
Error on data points is $<1.5 \%$.}
\end{figure}

Once $z$ is fixed, $\mathcal{I}_{av}(z)$ depends nonmonotonically
on $N$ and $L$; let us follow it for $N$ fixed and varying $L$ in
Fig.~\ref{fig:Info}. For $L$ small, due to the narrow distribution
discussed in the previous section, the value of the information is
high. When $L=\tilde{L}$, the population distribution on levels
reaches its extremal form and the information displays a minimum.
As $L$ increases, the information starts to rise again. So, the
main result is that, given a population number $N$, there is an
{\it optimal} lattice size $\tilde{L}$ for which the final
information is minimum; this value is typically intermediate
between the high-density and low-density regimes. The same happens
having fixed $L$ and letting $N$ vary: there is a minimum for
$N=\tilde{N}$, where $\tilde{N}$ depends on $L$.

This result implies that choosing an optimization strategy for the
spreading of information on the lattice is not trivial. Suppose
e.g. that we are given $N$ agents on a lattice and we want to
maximize the final average information $\mathcal{I}_{av}(z)$ by
varying the lattice size $L$ (starting from some $L_0$). This
optimization process is meant to be {\it local}: we are not
allowed to modify the size by several orders of magnitude, but
just around the starting size $L_0$. Then, the choice whether to
shrink or expand the lattice depends on $L_0$. If $L_0<\tilde{L}$,
increasing $L$ takes the system closer to the information minimum
($\mathcal{I}_{av}(z)$ decreases); decreasing $L$ increases
$\mathcal{I}_{av}(z)$ and is the right strategy. If on the other
hand $L_0>\tilde{L}$, increasing $L$ is the right strategy.

\begin{figure}[t]
\includegraphics[width=.5\textwidth]{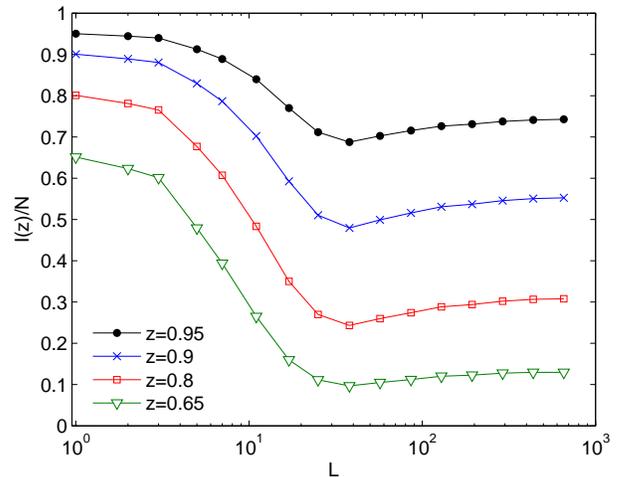}% Here is how to import EPS art
\caption{\label{fig:min_deg} (Color online) Semilog scale plot of
final degree of information per agent $\mathcal{I}_{av}(z)$ as a
function of the lattice size $L$, when $N=2^9$ (lines are guides to
the eye). Four different values of decay constant $z$ are
considered, as shown by the legend.}
\end{figure}

Fig.~\ref{fig:min_deg} shows that the depth of the information minimum
depends in turn on the decay constant $z$: as $z$ is
varied from $0$ to $1$, there are some curves (corresponding to
in-between values) which display a more emphasized minimum.

\begin{figure}
\includegraphics[height=64mm]{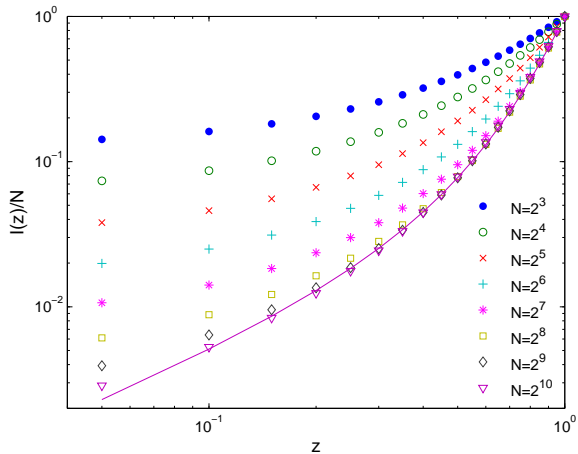}
\caption{\label{fig:pseudo_exp} (Color online) Final ($t=\tau$)
degree of information per agent $\mathcal{I}_{av}(z)$ versus the
decay constant $z$. The size of the lattice is fixed as $L=2^{4}$,
while several values of $N$ are considered and represented in
different colors and symbols. The curve depicted is the best fit
when $N  = 2^{10}$ ($\rho  > 1$), according to
Eq.~\ref{eq:info_hd}. Notice the existence of the fixed points
$z=0$, $\mathcal{I}_{av}(z) = \frac{1}{N}$ and $z=1,
\mathcal{I}_{av}(z)=1.$}
\end{figure}

Finally, in Fig.~\ref{fig:pseudo_exp} we show how the final
average degree of information $\mathcal{I}_{av}(z)$ depends on
$z$, for different values of $N$, once the size $L$ is fixed.
There are, as expected, two fixed points: when $z=1$ ($z=0$),
$\mathcal{I}_{av}(z)$ is equal to $1$ ($0$), irrespective of the
parameters ($N, L$) of the system. The function
$\mathcal{I}_{av}(z)$ cannot be determined but in two particular
regimes (low- and high-density).

When
$\rho=N/L^2$ is sufficiently low ($\rho < 2^{-8}$), the function
is well fitted by
\begin{equation}\label{eq:info_ld}
\mathcal{I}_{av}(z) = N^{z-1},
\end{equation}
within the error ($ < 3 \% $). When $\rho  > 1$,
$\mathcal{I}_{av}(z)$ is fitted by
\begin{equation}\label{eq:info_hd}
\mathcal{I}_{av}(z) =A\cdot z \rho \frac{(1-z^{B\cdot
L})^2}{(1-z)^2},
\end{equation}
with $A$, $B$ depending on $N$, $L$.

The two laws come from particular population distributions, as will be explained in the next section.

\section{\label{sec:Analytical} Analytical Results}
Consider a system with $N$ and $L$ fixed. Let $\mathcal{P}(t)$ be
the probability that at time $t$ an unaware agent is in contact
with at least 1 informed agent. Let $\mathcal{P}_l(k,s;\,t)$ be
the probability that at time $t$ an unaware agent is in contact
with $k+s$ informed agents, of which $k$ belonging to level $l$
and $s$ belonging to some other level. Then the evolution of the
system is governed by two master equations, one for the total
population:
\begin{equation}\label{eq:master1}
n(t+1) = n(t) + (N - n(t)) \, \mathcal{P}(t),
\end{equation}
and one for the level populations:
\begin{equation}\label{eq:master2}
n(l,\,t+1) = n(l,\,t) + (N-n(t)) \sum_{k,s} \mathcal{P}_{l-1}(k,s;\,t)
\frac{k}{k+s}.
\end{equation}
$\mathcal{P}(t)$ and $\mathcal{P}_l(k,s;\,t)$ are very complex functions
of their arguments and cannot be calculated in the general case. For
example, $\mathcal{P}(t)$ depends not only on the number of informed
agents $n(t)$ but also on their spatial distribution, hence on the instant
and the site where each of them has been informed (in other words, on the
history of the system). We will calculate the evolution of the system
in two particular cases, for high and low densities, and finally
compare the results with intermediate systems.

\begin{figure}
\includegraphics[width=0.48\textwidth]{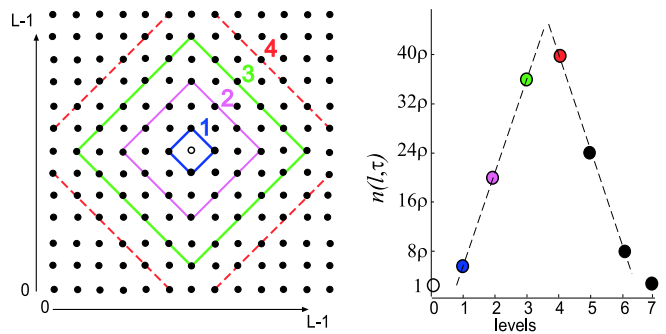}
\caption{\label{fig:reticolo} (Color online) Evolution of the system
in the high-density approximation for a lattice with $L=12$ and a
Source starting in the site with coordinates $(7,\,7)$. Left: the
wave front of information on the lattice at times $t=1$, 2, 3, 4 has
a square shape. Agents in the interior of the square are informed,
agents on the exterior are unaware. Times correspond to levels:
agents between the front at time 3 and that at time 4 belong to
level 4, and so on. Right: final distribution of agents on levels;
broken lines highlight the triangular shape of the distribution.}
\end{figure}

{\it High-density regime.} In this case ($\rho\rightarrow\infty$)
there are many agents on every site. If the agents on a site get
informed at a time $t$, we can suppose that at $t+1$ at least one
of them will jump on each of the four nearest- neighbor sites:
hence, all the unaware agents on the nearest- and
next-to-nearest-neighbor sites will get information at time $t+1$.
In this way (Fig. \ref{fig:reticolo}) information spreading among
agents amounts to propagation of information through the lattice.
A ``wave front'' of information travels with constant velocity: on
the interior sites are informed agents, on the exterior sites
unaware agents. If we suppose the Source to be at the center of
the lattice at $t=0$, at each instant the wave front is the locus
of points whose chemical distance from the center is $2t+1$.
Consequently, $n(t)=\rho(8t^2-4t+1)$, up to the half-filling time
$t_{out}\sim L/4$, when the front reaches the boundary of the
lattice; for $t>L/4$, the equation is
$n(t)=\rho(-8t^2+4t(2L+1)+(L+1)^2)$. The Population-Awareness Time
is $\tau\sim L/2$.

Almost all the agents on the wave front at time $t$ have received information at
time $t-1$: so, each new time step adds a new level, whose population
never changes at successive times. The population $n(l,\,t)$ is proportional to the length
of the wave front at the time $t=l$:
$n(l,\,t)\sim 4\rho(4l+1)$ up to $t=L/4$ and $n(l,\,t)\sim 4\rho(-4l+2L-1)$ up to $t=L/2$.
As can be seen from Fig. \ref{fig:reticolo}, the shape of the level distribution
at $t=\tau$ is triangular (compare this to the distribution for $L=16$ in
Fig. \ref{fig:level_distribution}).
The Final Information is proportional to $\rho$, according to the
formula
\begin{eqnarray}
\label{eq:info_hd_an} \mathcal{I}(z)&&= \sum_{l=0}^{N} n(l,\,\tau)
z^l \sim \sum_{l=0}^{L/4} 16\rho\,l z^l + \sum_{l=L/4+1}
^{\tau} 4\rho(2L-4l) z^l\nonumber\\
&& = 16z \rho \frac{(1-z^{L/4})^2}{(1-z)^2}.
\end{eqnarray}

A modified version of this equation, Eq.(\ref{eq:info_hd}), has
been used to fit the information curves for high-density regimes.

{\it Low-density regime.} In the case of low density ($\rho\ll 1$)
the time an informed agent walks before meeting an unaware agent
becomes very large. We can then assume that the agents between
each event have the time to redistribute randomly on the lattice,
that is, we adopt a mean-field approximation. Let $p=5/L^2$ be the
probability that two given agents, randomly positioned on the
lattice, are in contact (5 is the number of points contained in a
circle of radius 1). Hence, $(1-p)^{n(t)}$ is the probability for
an agent at time $t$ of not being in contact with any of the
$n(t)$ informed agents, and $\mathcal{P}(t)=1-(1-p)^{n(t)}$ is the
probability of being in contact with at least one informed agent.
Master equation (\ref{eq:master1}) becomes:
$$n(t+1) = n(t) + \left(N - n(t)\right) \left(1 - (1 - p)^{n(t)}\right),$$
and to first order in $p$:
\begin{equation}
n(t+1) = n(t) + p\,\left(N - n(t)\right)\,n(t).
\end{equation}

Thus, $n(t+1) = f(n(t))$: $f$ is a logistic-like map, with a
repelling fixed point in $0$ ($f'(0)=1+Np$), and an attracting
fixed point in $N$ ($f'(N)=1-Np$). Since $Np=5\rho\ll 1$, the
increment of $n(t)$ at each time step is very small (of order
$p$), and we can take the evolution to be continuous. The equation
becomes:
\begin{equation}
n(t+1)-n(t)\sim\frac{dn(t)}{dt}=p\,(N-n(t))\,n(t)
\end{equation}
and the solution, with the initial condition $n(0)=1$, is the sigmoidal function
\begin{equation}
n(t)=N\frac{e^{Npt}}{e^{Npt}+N-1}.
\end{equation}
The outbreak time, i.e. the flex of the curve, is in
$t_{out}=\frac{\mathrm{log}(N-1)}{Np}$, that is also the half-filling time, $n(t_{out})=N/2$.
The total population $N$ is reached only for $t=\infty$,
but we can take the PAT to be the time when $N-1$ agents have been informed:
\begin{equation}
\label{eq:ld_asymptotic}
\tau=\frac{2\,\mathrm{log}(N-1)}{Np}\sim\frac{2\,\mathrm{log}N}{Np}\sim\frac{\mathrm{log}N}{N}L^2,
\end{equation}
where the last result holds for $N$ large: hence, in the
low-density approximation the exponent for $L$ is $\beta=2$, while
the law for $N$ contains logarithmic corrections and the exponent
$\alpha$ cannot be defined. The first result in Eq.
(\ref{eq:ld_asymptotic}) shows that in this approximation
$\tau=2\,t_{out}$.

The quantity $\mathcal{P}_l(k,s;\,t)$ in Eq. (\ref{eq:master2}) is:

\begin{eqnarray*}
\mathcal{P}_l(k,s;\,t)= && \left(
\begin{array}{c}
n(l,t)\\
k
\end{array}
\right) \left(
\begin{array}{c}
n(t) - n(l,t)\\
s
\end{array}
\right) \\
&&\times p^{k+s}(1-p)^{n(t)-(k+s)}.
\end{eqnarray*}
The sum over $k$ and $s$ in Eq. (\ref{eq:master2}), using the
Chu-Vandermonde identity for binomial coefficients
\cite{vandermonde}, yields a master equation for the level
populations in the mean-field approximation:
$$n(l,\,t+1)=n(l,\,t)+(N-n(t))(1-(1-p)^{n(t)})\frac{n(l-1,\,t)}{n(t)},$$
and to first order in $p$:
$$n(l,\,t+1)=n(l,\,t)+p\,n(l-1,\,t)\,(N-n(t)).$$
Its continuous version is:
\begin{equation}\label{eq:highdil_lev_pop}
\frac{dn(l,\,t)}{dt}=p\,n(l-1,\,t)(N-n(t))
\end{equation}
that has to be solved for each $l$. For $l=1$, with the initial
condition $n(1,\,0)=0$, we get the solution
\begin{eqnarray}
n(1,\,t)&&=Np\,t-\mathrm{log}\left(e^{Np\,t}+N-1\right)+\mathrm{log}N\nonumber\\
&&=\mathrm{log}\left(n(t)\right).\nonumber
\end{eqnarray}
We then plug this solution into Eq. (\ref{eq:highdil_lev_pop}) to get $n(2,\,t)$, and so on.
It can be shown by induction that for every $l$, with the initial condition $n(l,\,t)=0$,
\begin{eqnarray}
n(l,\,t)&&=\frac{1}{l!}\left(Np\,t-\mathrm{log}\left(e^{Np\,t}+N-1\right)+\mathrm{log}N\right)^l\nonumber\\
&&=\frac{1}{l!}\left(n(1,\,t)\right)^l=\frac{1}{l!}\left[\mathrm{log}\left(n(t)\right)\right]^l.
\end{eqnarray}
This set of curves (not shown here) is similar to that of Fig. \ref{fig:levels},
with crossovers and different Outbreak Times.

The normalized level population at each $t$ is:
\begin{equation}
\frac{n(l,\,t)}{n(t)}=\frac{1}{n(t)}\frac{1}{l!}\left[\mathrm{log}\left(n(t)\right)\right]^l
=
\frac{e^{-\mathrm{log}\left(n(t)\right)}\left[\mathrm{log}\left(n(t)\right)\right]^l}{l!},
\end{equation}
hence, it is a Poisson distribution with mean $\mathrm{log}\left(n(t)\right)$.

The population distribution on levels at $t=\tau$ is
\begin{equation}
\label{eq:lev_distr_highdil}
n(l,\,\tau)=\frac{\left(\mathrm{log}N\right)^l}{l!},
\end{equation}
independent of $p$ (hence of $L$). A modified version of this
distribution, Eq. (\ref{eq:fitting_function}), has been used to
fit the numerical curves.

The total information is
\begin{eqnarray}
\mathcal{I}(t,\,z)&&=\sum_{l=0}^{N}n(l,\,t)z^l=\sum_{l=0}^{N}\frac{1}{l!}
\left[\mathrm{log}\left(n(t)\right)\cdot z\right]^l\sim\nonumber\\
&&\sim e^{\mathrm{log}\left(n(t)\right)\cdot z}=n(t)^z.
\end{eqnarray}
In particular, $\mathcal{I}(\tau,\,z)=N^z$, in agreement with Eq. (\ref{eq:info_ld}).

\begin{figure}
    \includegraphics[height=75mm]{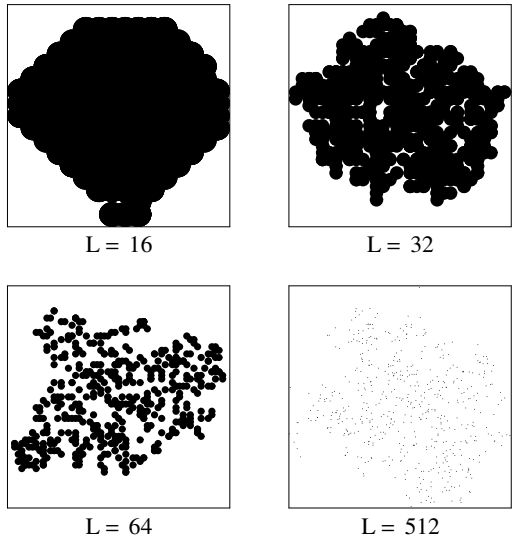}
    \caption{\label{fig:frames} Snapshots of four systems with $N=1024$ and $L=16$, 32, 64, 512, all at an instant near to
    the half-filling time. Only informed agents are shown; they are represented as circles of radius 1. Notice
    that the high-density picture of a connected set of informed agents holds up to $L=\tilde{L}=64$. For $L\geq 64$,
    the picture breaks down and the system is better described by a low-density approximation ($L=512$).}
\end{figure}

In conclusion, we have examined the system in two different
regimes, both optimal for information spreading. The worst case
for information spreading, at $\tilde{L}$, seems to correspond to
crossover between these two regimes, as shown in
Fig. \ref{fig:frames}.\\

\section{\label{sec:Conclusions} Conclusions and perspectives}
We have presented a model of information spreading amongst
diffusing agents. The model takes into account a population made
up of agents who are socially, as well as geographically, dynamic.
Moreover, it allows for possible alteration of information
occurring during the transmission process, by introducing a decay
constant $z$.

Investigations are lead both by means of numerical simulations and
of analytical methods valid in the
high- and low-density regimes.

The main results are two. First: the time it takes the piece of
information to reach the whole population of $N$ agents, distributed
on a lattice sized $L$, depends on $N$ and $L$ according to a power
law. This behavior holds over a wide range, where exponents are
found to be constant and noninteger. Second: the final ($t=\tau$)
average degree of information $\mathcal{I}_{av}(z)$ for a fixed
population $N$ (lattice size $L$) shows a surprisingly non-monotonic
dependence on the lattice size $L$ (on the population $N$), with the
occurrence of a minimum. This means that, from an applied
perspective, an optimization strategy for $\mathcal{I}_{av}(z)$ is
possible with respect to $N$ and $L$.

Extensions of our model to networks embedded in topologically
different spaces are under study.

\end{document}